\documentclass[journal=nalefd,manuscript=letter]{achemso}
\usepackage{color}
\usepackage{lipsum,amsmath}
\usepackage{pifont}   
\usepackage{graphicx} 
\usepackage{dcolumn}  
\usepackage{bm}       
\usepackage{amsfonts} 
\usepackage{amssymb}  
\usepackage{multirow} 
\usepackage{natbib}
\usepackage{multicol}
\usepackage{tikz}
\usepackage{siunitx}
\usepackage{placeins}
\usepackage{braket}
\usepackage{nicefrac}
\usepackage{bm}
\usepackage{physics}
\usepackage{grffile}

    \newcommand{\m}[1]{\bm{\mathsf{#1}}}

\usepackage[english]{babel}
\usepackage[autostyle, english = american]{csquotes}
\MakeOuterQuote{"}

\title{Transient spin injection efficiencies at ferromagnet/metal interfaces}

\author{P. Elliott}
\affiliation{Max-Born-Institut  f\"ur Nichtlineare Optik und Kurzzeitspektroskopie, Max-Born-Strasse 2A, 12489 Berlin, Germany}
\author{A. Eschenlohr}
\affiliation{Faculty of Physics and Center for Nanointegration Duisburg-Essen (CENIDE), University Duisburg-Essen, Lotharstr. 1, 47057 Duisburg, Germany}
\email{andrea.eschenlohr@uni-due.de}
\author{J. Chen}
\affiliation{Faculty of Physics and Center for Nanointegration Duisburg-Essen (CENIDE), University Duisburg-Essen, Lotharstr. 1, 47057 Duisburg, Germany}
\author{S. Shallcross}
\affiliation{Max-Born-Institut  f\"ur Nichtlineare Optik und Kurzzeitspektroskopie, Max-Born-Strasse 2A, 12489 Berlin, Germany}
\author{U. Bovensiepen}
\affiliation{Faculty of Physics and Center for Nanointegration Duisburg-Essen (CENIDE), University Duisburg-Essen, Lotharstr. 1, 47057 Duisburg, Germany}
\author{J.~K. Dewhurst}
\affiliation{Max-Planck-Institut f\"ur Mikrostrcukturphysik, Weinberg 2, 06120 Halle, Germany}
\author{S. Sharma}
\affiliation{Max-Born-Institut  f\"ur Nichtlineare Optik und Kurzzeitspektroskopie, Max-Born-Strasse 2A, 12489 Berlin, Germany}

\begin{document}

\date{\today}

\begin{abstract} 

Spin injection across interfaces driven by ultrashort optical pulses on femtosecond timescales constitutes a new way to design spintronics applications. Targeted utilization of this phenomenon requires knowledge of the efficiency of non-equilibrium spin injection. From a quantitative comparison of \textit{ab-initio} time-dependent density functional theory and interface-sensitive, time-resolved non-linear optical experiment, we determine the spin injection efficiencies (SIE) across ferromagnetic/metal interfaces and discuss their microscopic origin, i.e. the influence of spin-orbit coupling and the interface electronic structure. Moreover, we find that the SIE can be optimized through laser pulse and materials parameters, namely the fluence, pulse duration, and substrate material. 

\end{abstract}

\maketitle

\section{Introduction}

Spin injection is at the heart of a potential new electronics based on spintronics \cite{Hirohata2020} and valleytronic\cite{Jin2018} devices. Device applications require optically driven spin injection, a phenomenon already realised in metallic heterostructures on femtosecond (fs) timescales \cite{malinowski2008, Battiato2010, melnikov2011, Eschenlohr2013, turgut2013, kampfrath2013, eschenlohr2020}. For ultrafast spintronic concepts, the appropriate structure consists of a non-ferromagnetic metal in contact with a ferromagnetic metal. Light induced excitation of the ferromagnet causes a spin-selective excitation due to the exchange-split band structure, with the flow of spin current across the interface leading to spin injection into the non-ferromagnetic metal. However, not all of the spins excited in the ferromagnetic layer are transferred into the non-ferromagnetic counterpart due to several mechanisms, chiefly charge carrier spin-flip scattering. For a realistic utilisation of these ultrafast spin currents, knowledge of the efficiency of this fs spin injection and the microscopic mechanisms behind it are necessary, in particular in order to understand any fundamental limitations and to allow for optimisation of the spin injection process. 

The efficiency of spin injection can be estimated by measuring the ratio of the amount of spin-moment lost in the ferromagnetic material and the amount gained by the non-ferromagnet. Experimental estimates of the spin injection efficiency (SIE) from a transition metal ferromagnet into a non-ferromagnetic metal with fs time resolution have been reported for selected interfaces: For polycrystalline Ni/Au interfaces, a SIE of about 50\% at $\leq500$~fs has been inferred from the decrease of the magnetization in the ferromagnetic layer as compared to the induced magneto-optical signal in the non-ferromagnetic substrate, with the help of model calculations assuming superdiffusive spin transport \cite{hofherr2017}. In our previous study of 3-5 monolayers of epitaxial Co on Cu(001), we found a SIE of 25-40\% at 35~fs after optical excitation via comparison of interface-sensitive magneto-optical measurements of the Co demagnetization and \textit{ab-initio} calculations of the transiently induced Cu spin moment \cite{chen2019}. In this case we found OISTR\cite{dewhurst2018} to be the dominating mechanism for spin transfer across the interface. Furthermore, the combination of \textit{ab-initio} quantum transport calculations and interface-sensitive magneto-optical experiments on epitaxial Au/Fe/MgO(001) heterostructures allowed for an estimate of 70\% of the spins within a 250~fs long spin current pulse transferring torque at the Au/Fe interface \cite{alekhin2017}. 

So far, no systematic and material-dependent understanding of ultrafast spin injection has been achieved, owing to the experimental difficulty and theoretical complexity of this process. Especially the influence of the interface electronic structure and morphology as well as the spin-orbit coupling (SOC) strength should be considered in a proper description of the non-equilibrium state. Moreover, the details of the optical excitation, namely the photon energy, duration, and fluence of the driving laser pulse, can be expected to determine the fundamental light-induced processes underlying the SIE. 

This motivates the use of \textit{ab-initio} theory, namely time-dependent density functional theory (TDDFT), for a quantitative determination and prediction of the transient SIE. In particular, this DFT-based approach is material-specific, and has been demonstrated to deliver an analysis in quantitative agreement with experiments for spin dynamics within the first $\approx100$~fs \cite{chen2019,Siegrist2019,hofherr2020,dewhurst2020,sharma22,dewhurst2018a}. This focus on the initial fs dynamics determined by the optical excitation also promises to make tuning of the transient spin injection with the material combination and/or pulse parameters possible. With modern methods of optical pulse shaping such control might in the future be considered, with the goal of designing specific spin current pulses for ultrafast spintronics applications, such as spin-transfer torque (STT) devices \cite{SLONCZEWSKI1996, Berger1996} where such spin current injection may be used to control the excitation of the magnetic moment.

In the present work, we use epitaxial layers of Co/X(001) [X= Al, Cu, Au and Pt] interfaces to showcase the optically induced ultrafast demagnetization of Co leading to non-equilibrium spin injection into X. Starting from a quantitative agreement between fs time-resolved experiment and \textit{ab-initio} theory, we subsequently investigate the transient spin injection efficiency (SIE) and study how the spin injection changes as a function of time upon laser excitation. This allows us to analyze theoretically the microscopic physical processes limiting the spin injection on fs timescales. We find that the interplay of available states and SOC strength of the non-ferromagnetic metal plays a dominating role in the reduction of the transient spin injection efficiency, from very high values that are attained at early times. Moreover, we demonstrate that the pump laser pulse parameters play a significant role in controlling the fs spin injection, and discuss ways to optimize the SIE for future ultrafast spintronics applications. 

\section{Theoretical Methodology}

To calculate transient magnetic spin-moments in laser pumped materials we have used the fully {\it ab-initio} state-of-the-art time dependent density functional theory (TDDFT)\cite{RG1984,my-book}, that rigorously maps the computationally intractable problem of interacting electrons to the Kohn-Sham (KS) system of non-interacting fermions in a fictitious potential, a problem that can be solved by modern computing clusters. The time-dependent KS equation is:

\begin{align}
i &\frac{\partial \psi_{j{\bf k}}({\bf r},t)}{\partial t} =
\Bigg[
\frac{1}{2}\left(-i{\nabla} +\frac{1}{c}{\bf A}_{\rm ext}(t)\right)^2 +v_{s}({\bf r},t) \nonumber \\
& + \frac{1}{2c} {\sigma}\cdot{\bf B}_{s}({\bf r},t) + \frac{1}{4c^2} {\sigma}\cdot ({\nabla}v_{s}({\bf r},t) \times -i{\nabla})\Bigg]
\psi_{j{\bf k}}({\bf r},t)
\label{e:TDKS}
\end{align}
where ${\bf A}_{\rm ext}(t)$ is a vector potential representing the applied laser field. It is assumed that the wavelength of the applied laser is much greater than the size of a unit cell and the dipole approximation can be used, i.e. the spatial dependence of the vector potential is disregarded. This constitutes a reasonable assumption for the laser wavelengths in the near-infrared range used here. The KS potential $v_{s}({\bf r},t) = v_{\rm ext}({\bf r},t)+v_{\rm H}({\bf r},t)+v_{\rm xc}({\bf r},t)$ is decomposed into the external potential $v_{\rm ext}$, the classical electrostatic Hartree potential $v_{\rm H}$ and the exchange-correlation (XC) potential $v_{\rm xc}$. Similarly, the KS magnetic field is written as ${\bf B}_{s}({\bf r},t)={\bf B}_{\rm ext}(t)+{\bf B}_{\rm xc}({\bf r},t)$ where ${\bf B}_{\rm ext}(t)$ is an external magnetic field and ${\bf B}_{\rm xc}({\bf r},t)$ is the exchange-correlation (XC) magnetic field.  In the present work we have used adiabatic local density approximation for XC potential. ${\sigma}$ are the Pauli matrices and the final term of Eq.~\eqref{e:TDKS} is the spin-orbit coupling term. In the fully non-collinear spin-dependent version of this theory\cite{krieger2015,dewhurst2016} the orbitals $\psi$ are two component Pauli spinors and from these the magnetisation density can be calculated as:

\begin{eqnarray}\label{st}
{\bf m}({\bf r},t)=\sum_j \psi^*_j({\bf r},t){\m \sigma}\psi_j({\bf r},t),
\end{eqnarray}
The integral of this vector field over space gives the total magnetic moment as a function of time (M(t)).
All calculations are performed using the highly accurate full potential linearized augmented-plane-wave method\cite{singh}, as implemented in the Elk \cite{elk,dewhurst2016} code.
The Brillouin zone was sampled with a $15\times 15\times 1$ k-point mesh. For time propagation the algorithm detailed in Ref.~\cite{dewhurst2016} was used with a time-step of $2.42$ atto-seconds. The final magnetisation value converges with the above mentioned computational parameters to 1.73~$\mu_{\rm B}$ per Co atom for the bulk material. We restrict ourselves to a few hundred fs regime as the non-inclusion of nuclear dynamics prevents us from describing physics at longer time scales. To ensure that the theoretical calculation provides a good representation of the actual experimental situation, which is introduced in more detail in the following section, we have performed a calculation for the experimentally studied material, Co/Cu(001), using the experimental laser pulse parameters (pulse duration 35~fs, wavelength 800~nm/frequency 1.55~eV and absorbed pump fluence of 0.25~mJ/cm$^2$).

\section{Experimental Methodology}

In order to achieve direct experimental access to the fs spin dynamics at a ferromagnet/metal interface, we investigate ultrathin epitaxial Co films with 3 monolayer (ML) thickness on a Cu(001) surface. After the Cu(001) single crystal substrate is prepared by several sputtering-annealing cycles, Co is deposited by electron beam evaporation. These Co/Cu(001) films have atomically sharp interfaces \cite{weber1996,jaehnke1999}, making  them an ideal model system for the comparison to \textit{ab-initio} theory. Preparation, characterization, and the subsequent time-resolved measurements are performed \textit{in situ} at room temperature in ultrahigh vacuum at a pressure of $<10^{-10}$~mbar. 

An interface-sensitive probe, namely second harmonic generation (SHG) \cite{guedde1999,chen2017} is employed. In centrosymmetric crystals such as that investigated here, SHG is only generated where spatial inversion symmetry is broken, i.e. at interfaces. This optical technique, which employs ultrashort laser pulses, moreover enables efficient time-domain analysis with fs time resolution. We conduct a pump-probe experiment with near-infrared laser pulses with 800~nm wavelength and a pulse duration of 35~fs (full width half maximum), generated with a cavity-dumped Ti:Sapphire oscillator. After excitation with s-polarized pump pulses of an incident pump fluence of $4\pm2$~mJ/cm$^2$, the resulting spin dynamics at the Co/Cu(001) interface are probed with SHG at 400~nm wavelength. 
This incident fluence corresponds to the absorbed fluence of 0.25~mJ/cm$^2$ that is used in the TDDFT calculations for this interface, ensuring that the excitation conditions are the same in experiment and theory. 
The Co films are magnetized in the direction parallel to the sample surface and perpendicular to the optical plane by an external magnetic field. We thus detect SHG of the p-polarized probe pulse in transversal geometry, by means of single photon counting after filtering with a BG39 filter and monochromatization with a grating monochromator, as shown schematically in Fig.~\ref{exp-th}(a). The slightly non-collinear incidence of the pump and probe beams ensures that the probe signal can be reliably spatially separated from the reflected pump beam and the pump-probe cross-correlation. 

We acquire the second harmonic (SH) intensities $I^{\uparrow,\downarrow}$ for opposite orientations of the magnetization \textbf{M} of the Co film, from which we derive the SH fields, $|E^{2\omega}_{\mathrm{even}}| \approx \sqrt{\frac{I^{\uparrow}+I^{\downarrow}}{2}}$ and $|E^{2\omega}_{\mathrm{odd}}| \approx \frac{I^{\uparrow}-I^{\downarrow}}{4|E^{2\omega}_{\mathrm{even}}|}$. These fields are even and odd with respect to a sign change of \textbf{M}, respectively, and are thus considered magnetization-\textit{in}dependent and -dependent for $E^{2\omega}_{\mathrm{even}} >> E^{2\omega}_{\mathrm{odd}}$, as is the case for Co/Cu(001) \cite{guedde1999, conrad2001}. The time-dependent changes of $E^{2\omega}_{\mathrm{even,odd}}$ are then normalized to their respective values before optical excitation, resulting in the time-dependent relative changes $\Delta^{2\omega}_{\mathrm{even,odd}}=\frac{E^{2\omega}_{\mathrm{even,odd}}(t)}{E^{2\omega}_{\mathrm{even,odd}}(t<0)}-1$. In this study, we focus on the magnetization-dependent observable $\Delta^{2\omega}_{\mathrm{odd}}$, which provides information on the ultrafast spin dynamics in the Co film. We note that while transient spin injection has been probed in Au with SHG \cite{melnikov2011}, we are not sensitive to the transiently induced spin polarization in the non-ferromagnetic substrate here, since the much lower spin-orbit coupling of Cu compared to Au diminishes the respective magneto-optical response. Further details on our SHG setup as well as the charge dynamics measured through $\Delta^{2\omega}_{\mathrm{even}}$ are provided in Refs.~\cite{chen2017,chen2019}. 

\section{Results}

\begin{figure}[ht]
\includegraphics[width=0.4\columnwidth, clip]{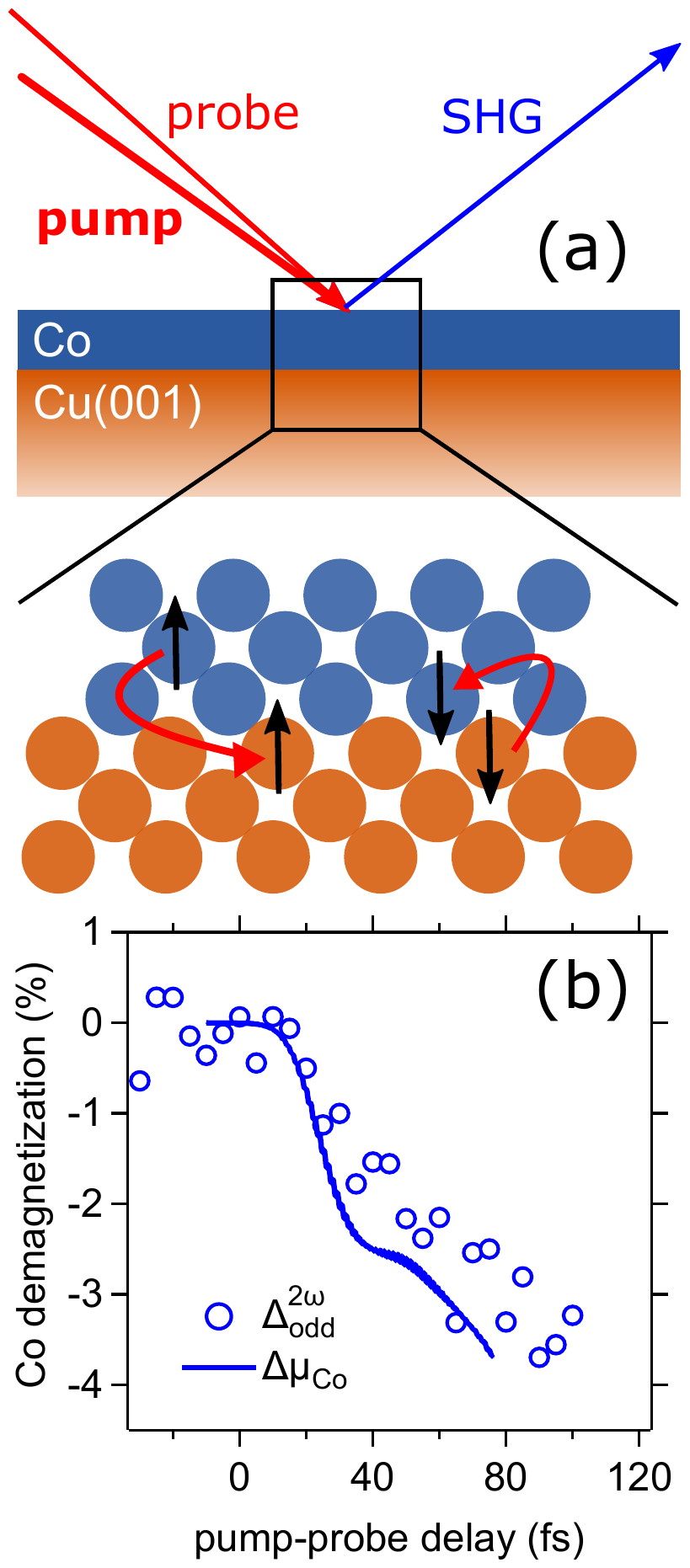} 
\caption{(a) Sketch of the experimental geometry and microscopic processes at Co/Cu(001) interface: The ultrafast demagnetization of Co after excitation with a near-infrared pump pulse is probed by second harmonic generation (SHG). \textit{Ab-initio} TDDFT reveals that spin transfer across the interface involves majority spin injection from Co to the Cu substrate as well as minority spin back-transfer from Cu to Co. (b) Comparison of experimental observation of ultrafast demagnetization of Co via the relative change of magnetization-dependent SHG signal $\Delta^{2\omega}_{\mathrm{even}}$ (blue circles) with theoretical calculation of relative change in Co spin moment $\Delta\mu_{\mathrm{Co}}$ after laser excitation (blue line).} 
\label{exp-th}
\end{figure}
Fig.~\ref{exp-th}(b) shows the experimentally measured ultrafast demagnetization $\Delta^{2\omega}_{\mathrm{odd}}$ of the 3~ML thick Co film at the Co/Cu(001) interface, depending on the pump-probe time delay. We observe a relative change of about 4\% within the first 100~fs. Since we are here interested in the non-equilibrium spin-dependent processes only, we focus on this initial demagnetization, after which relaxation to the ground state through electron-phonon scattering within several picoseconds sets in \cite{chen2017, chen2019}. We achieve microscopic insight into the spin dynamics underlying the optically induced demagnetization of Co by a direct comparison with the relative change of the magnetic moment of a 3~ML thick Co layer on top of a 7~ML thick Cu(001) substrate calculated with TDDFT for the same excitation conditions, i.e. absorbed pump fluence. We note that time zero corresponds to the peak of the envelope of the pump-probe cross-correlation in experiment respectively to the peak of the pump pulse envelope in theory. The quantitative agreement between experiment and \textit{ab-initio} theory observed in Fig.~\ref{exp-th}(b) allows us to proceed to an investigation of the microscopic mechanisms via TDDFT.

The Co demagnetization results from the interplay of two processes, namely spin transfer across the Co/Cu(001) interface and spin flips mediated by spin-orbit coupling. Roughly during the first 35~fs, while the pump pulse is present, spin transfer dominates the dynamics. It proceeds via injection of majority spin electrons from Co to Cu, and backtransfer of minority spin electrons from Cu to Co, see Fig.~\ref{exp-th}(a), as found in our earlier study \cite{chen2019}. We have previously explained this transient spin injection through the so-called OISTR mechanism, i.e. optically induced intersite spin transfer \cite{dewhurst2018,dewhurst2018a}, which depends on the details of the density of states (DOS) at the interface as well as the optical pulse parameters, and thus allows the tuning of the spin transfer and attendant SIE, as will be shown later. 

The spin transfer is suppressed through SOC, which leads to spin flips and thus further demagnetization after 35~fs, but at the cost of mixing of the majority and minority spin channels. Therefore, it can be expected that the transient SIE results from a competition of spin injection, i.e. OISTR, and SOC-mediated spin flips. We analyze this interplay in detail in the following, for a slightly shorter pump pulse that allow us to conserve computation time while still remaining in an excitation regime that corresponds to a realistic experimental situation.

\begin{figure}[t!]
\includegraphics[width=\columnwidth, clip]{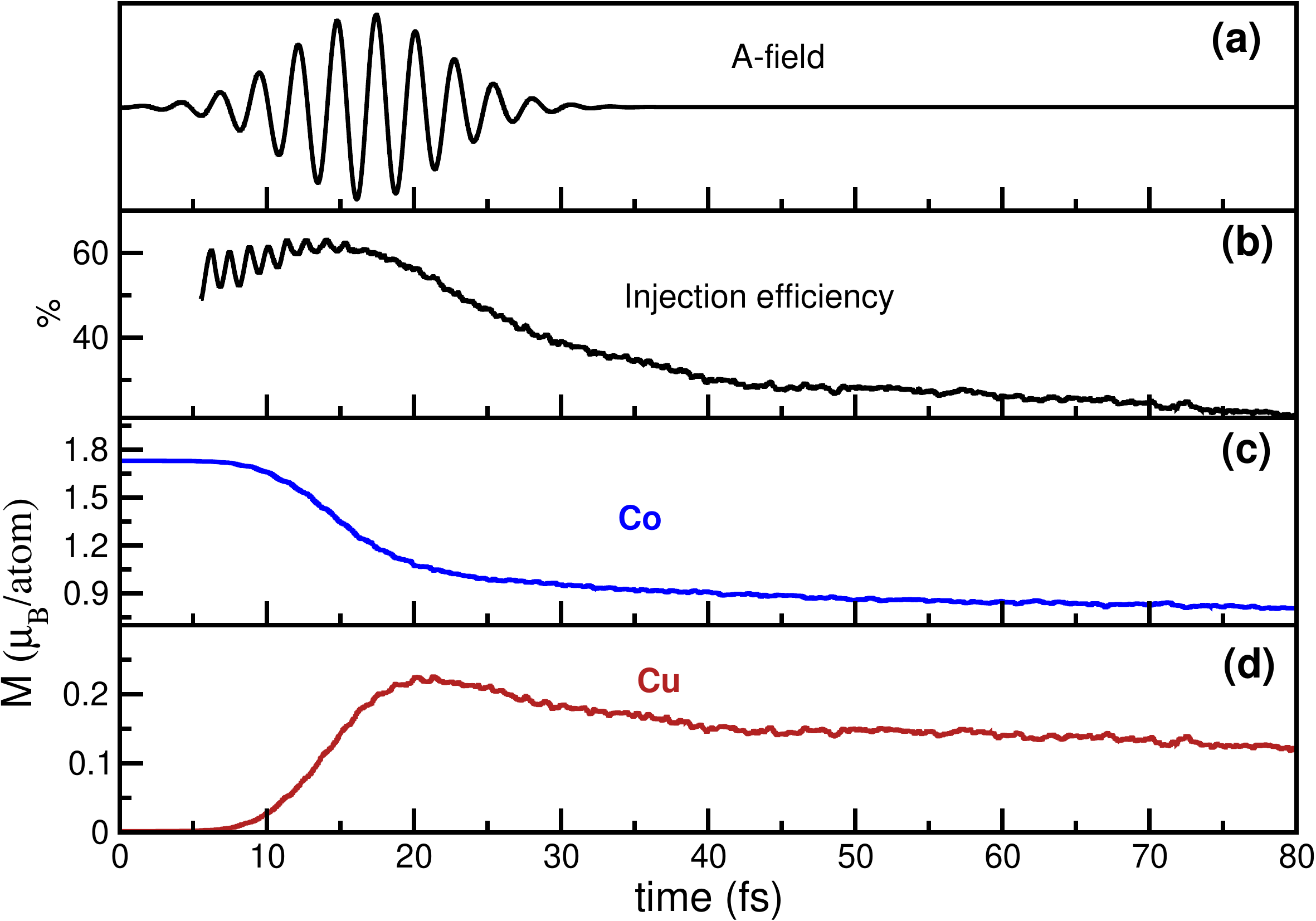}
\caption{(a) Vector potential of the pump pulse with central frequency of 1.55~eV, FWHM of 12~fs and incident fluence of 15~mJ/cm$^2$, (b) spin injection efficiency (in percentage) (c) magnetic moment per Co atom (in $\mu_B$), (d) magnetic moment per Cu atom (in $\mu_B$) in 3Co/5Cu. } \label{co3cu5-s}
\end{figure}
In Fig.~\ref{co3cu5-s}, we present the transient spin injection efficiency for 3 layers of Co on top of 5 layers of Cu(100) following excitation by an ultrafast laser pulse, whose vector potential is shown in Fig.~\ref{co3cu5-s}(a). 
The transient efficiency of this injection, as a percentage, can be defined by:

\begin{equation}
\label{e:injdef}
\text{Spin injection efficiency} = \frac{\Delta M_{NM} (t)}{\Delta M_{FM} (t)} \times 100
\end{equation}
where $\Delta M_{NM} (t),\Delta M_{FM} (t)$ are the transient changes in atomic magnetic moments from the ground-state for the normal metal (NM) and ferromagnet (FM), respectively. Due to spin-orbit coupling (SOC), the magnetic moment is not a conserved quantity, and so the efficiency varies on ultrafast time scales. 

\begin{figure}[t!]
\includegraphics[width=\columnwidth, clip]{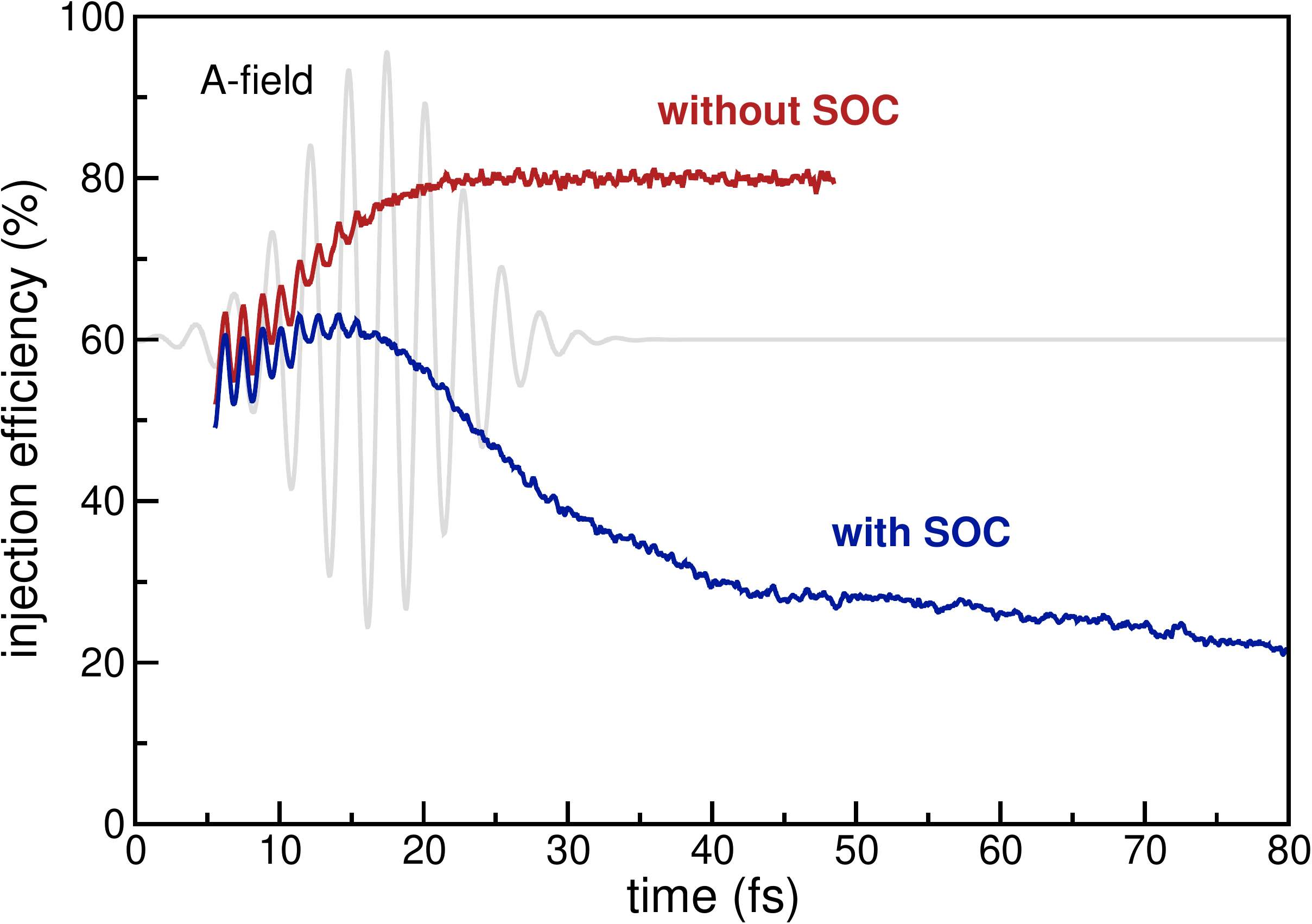}
\caption{Transient injection efficiency (given as a percentage, see Eq.~\ref{e:injdef}) for a 3Co/5Cu interface pumped with a laser pulse with a central frequency of 1.55~eV, a FWHM of 12~fs, and an incident fluence of 15~mJ/cm$^2$. The results are shown for calculations with and without spin-orbit coupling (SOC). Evidently, SOC plays a crucial role in degrading the injection efficiency at longer times. Note that without SOC the transient injection efficiency saturates below 100\% due to moment residing in delocalized interstitial states.} \label{soc}
\end{figure}

The transient SIE is shown in Fig.~\ref{co3cu5-s}(b), where it is initially quite high, $\approx 60\%$ before falling to a constant value of 20\% over a period of $20$ fs. The rising envelope of the pump pulse, where the number of photons increases, leads to more and more carriers being excited, resulting in an increase in the SIE in this initial period.
We can understand the drop in the moment and SIE from the time-dependent average atomic moments of Co and Cu, as shown in Figs.~\ref{co3cu5-s}(c) and \ref{co3cu5-s}(d) respectively. Initially, the drop in the Co moment is accompanied by a corresponding rise in the Cu moment explaining the high efficiency at early times. Following this, the moments on both Co and Cu begin demagnetizing with the Co moment losing a significant amount. The net result is that there is less moment available for injection from the Co and less injected moment that survives in the Cu, both of which, from Eq.~(\ref{e:injdef}), cause the spin injection efficiency to drop. Eventually, this demagnetization process saturates and slows, and the efficiency levels off at its final value of just $20\%$. 

The demagnetization process observed in both Co and Cu is due to spin-orbit mediated spin-flips, first predicted in Ref.~\cite{KDES15}. This can be seen in Fig.~\ref{soc}, where the TDDFT simulation is performed with and without the SOC term in the Hamiltonian, Eq.~(\ref{e:TDKS}). Without SOC, the injection efficiency begins at $60\%$, the same as with the SOC term, but increases during the pulse before reaching a constant value of $80\%$ when the pulse has ended. Thus it is clear that SOC is responsible of the demagnetization of Co and Cu and the ensuing loss of efficiency.  

In the absence of SOC the total moment is a conserved quantity, and thus we would expect the efficiency to be $100\%$, as any moment lost in the Co must be transferred to the Cu in order to conserve the total. However, the efficiency defined in Eq.~(\ref{e:injdef}) employs the atomic moments, defined as the integral of the magnetization density in a spherical volume centred on each atom. The spin density that falls outside these spheres is known as the interstitial moment and is delocalized in space. These delocalized states may indicate spin currents that would transport moment away from the laser excited region, however for practical reasons, our simulations, like the vast majority of DFT simulations, are performed with periodic boundary conditions. Thus we cannot observe this transport, only the excitation to current carrying states. The efficiency defined in Eq.~(\ref{e:injdef}) therefore measures the spin injection that will remain in the Cu substrate.  

\begin{figure}[t!]
\includegraphics[width=\columnwidth, clip]{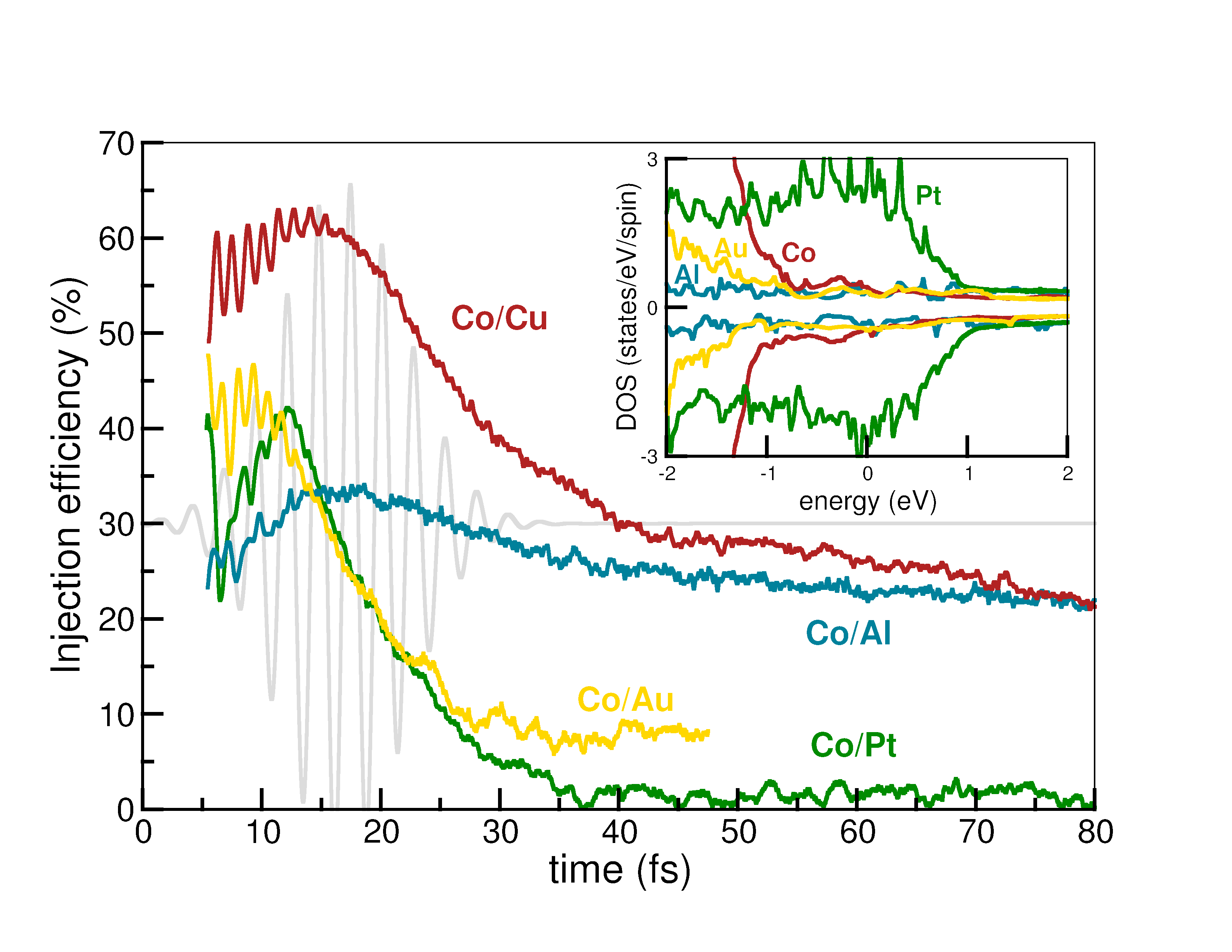}
\caption{Transient injection efficiency (given as a percentage, see Eq.~\ref{e:injdef}) for 3Co/5X interfaces (X= Cu, Al, Pt, Au) pumped with a laser pulse of central frequency 1.55~eV, FWHM 12~fs, and incident fluence 15~mJ/cm$^2$. This highlights the effect of the substrate on injection efficiency which decreases strongly as a function of time for the high SOC substrates (Pt, Au). The inset shows the majority spin (positive axis) and minority spin (negative axis) projected density of states (DOS) for Al ($sp$-orbitals) and Cu, Pt, Au ($d$-orbtials). The Fermi energy has been shifted to $0$ eV in each case.} \label{subs}
\end{figure}

Having identified the laser pulse and SOC as responsible for the transient behavior of the spin injection efficiency, it is natural to ask how we may optimize and control this behavior. In Fig.~\ref{subs}, we demonstrate the importance of the substrate material from Cu to Al, Pt, and Au on the initial and final SIE. 

The initial SIE is governed by the laser excitation of electrons from occupied to unoccupied states. By changing the substrate material, we can change the energy, character, and density of states available in the substrate and thus affect this excitation. For example, while both Cu and Au have almost fully occupied $d$-bands below the Fermi level, the bandwidth of Cu is lower than Au (see insert of Fig.~\ref{subs}). Thus, the density of states near the Fermi level is much higher for Cu than Au, meaning in the same frequency window, there are more occupied states available for excitation in Cu than Au, explaining why the initial SIE is higher for Co/Cu than Co/Au. The DOS of Pt is similar to Au except shifted in energy so that there are $d$-states above the Fermi level, as Pt has fewer valence electrons than Au (also shown in the inset of Fig.~\ref{subs}). Hence the similar behavior of Pt to Au in Fig.~\ref{subs} during the initial time (in this case, up to the time of the laser pulse peak). For Al, the states around the Fermi level have a completely different character than Cu, Pt, or Au, as they are mainly hybrid $s$- and $p$-states and the DOS is much smaller due to the lower number of valence electrons. Consequently, Al has the lower injection efficient during the initial excitation before SOC becomes active.

\begin{figure}[t!]
\includegraphics[width=\columnwidth, clip]{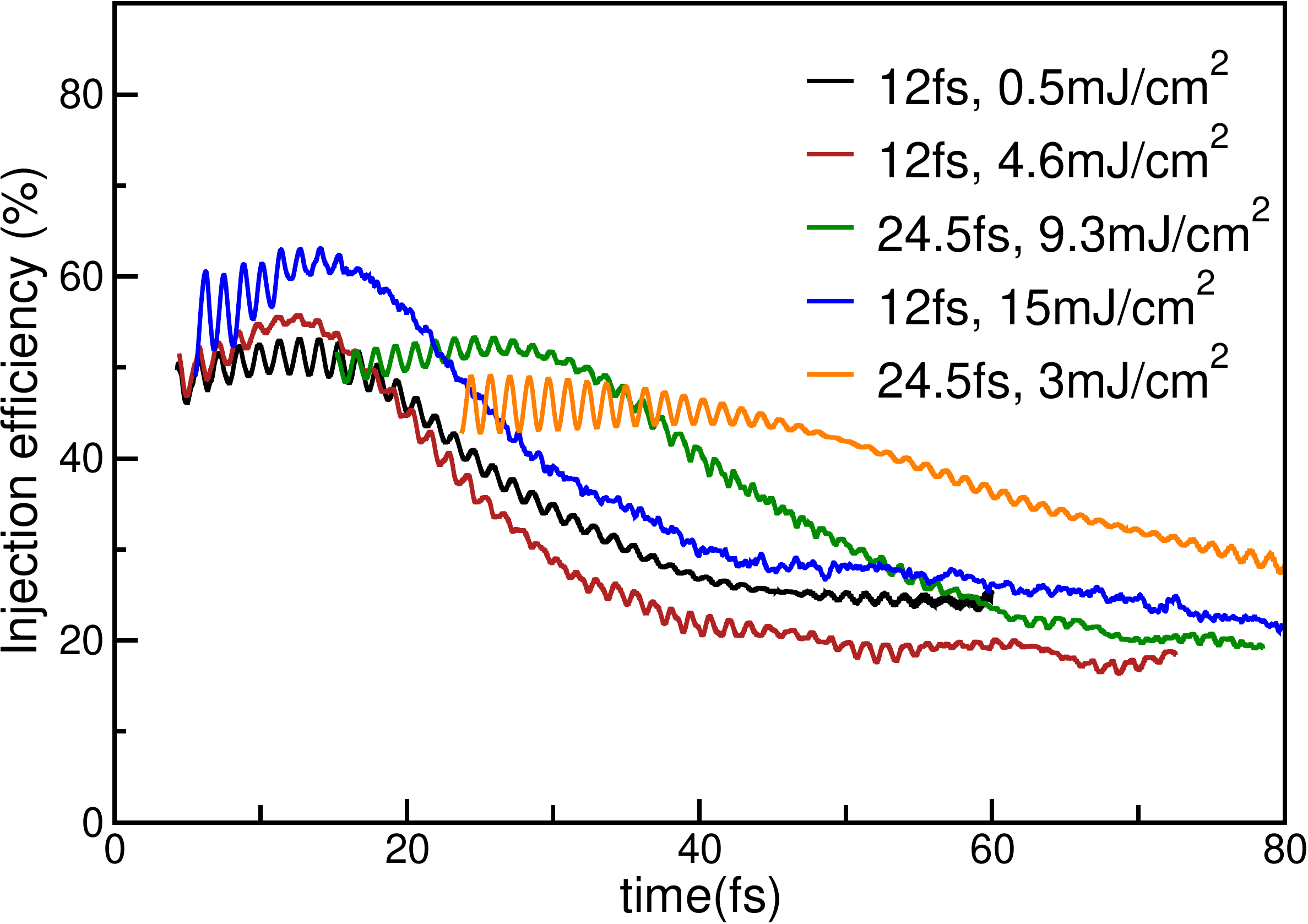}
\caption{Transient injection efficiency (given as a percentage) for a 3Co/5Cu interface pumped with various pulses (see legend for pulse parameters).} \label{pulse}
\end{figure}

For later times in Fig.~\ref{subs}, the transient behavior of the SIE is markedly different for the different substrates. While the SIE for Co/Cu decays to a constant, it decays to almost zero for Au and actually reaches zero for Pt. This confirms the observation that SOC is responsible for this decay as the SOC strength is known to be higher in Au and Pt than Cu. For a material like Al where SOC is not strong, there is much less decay of the SIE (there is still some decay due to the SOC demagnetization in the Co layers). 

The above results were calculated for a particular laser pulse, however the spin-orbit driven spin-flips and spin injection are non-linear phenomena that depend on how strongly the system is perturbed from equilibrium, and so the laser pulse parameters may also be tuned to control the SIE. In Fig.~\ref{pulse}, the effect of varying the duration (full width at half maximum, FWHM) and fluence of the laser pulse on the SIE is shown. Keeping the FWHM fixed at $12$~fs while varying the fluence demonstrates the non-linearity present in the problem. Increasing the fluence can cause both the non-linearity in the initial excitation as well as the subsequent SOC demagnetization. Increasing the fluence from $0.5$~mJ/cm$^2$ to  $4.6$~mJ/cm$^2$ will increase the total number of electrons excited but this causes greater demagnetization leading to a larger decay in the SIE for later times. Increasing further to $15$ mJ/cm$^2$, we see the SIE at initial times increases due to non-linear optical transitions, however the decay due to SOC is also larger with the end result being a SIE similar to the $0.5$ mJ/cm$^2$ case. Increasing the FHWM to $24.5$~fs does not affect the initial SIE which remains around $50\%$ at the center time of all pulse. Increasing the fluence from $3$~mJ/cm$^2$ to $9.3$~mJ/cm$^2$ for this longer duration pulse leads to stronger demagnetization and a larger decay of the SIE, as was the case for the $12$~fs pulses, although the decay time is increased. Thus, to have a larger SIE for a longer time, the best combination is a longer laser pulse with a smaller fluence.

\begin{figure}[t!]
\includegraphics[width=\columnwidth, clip]{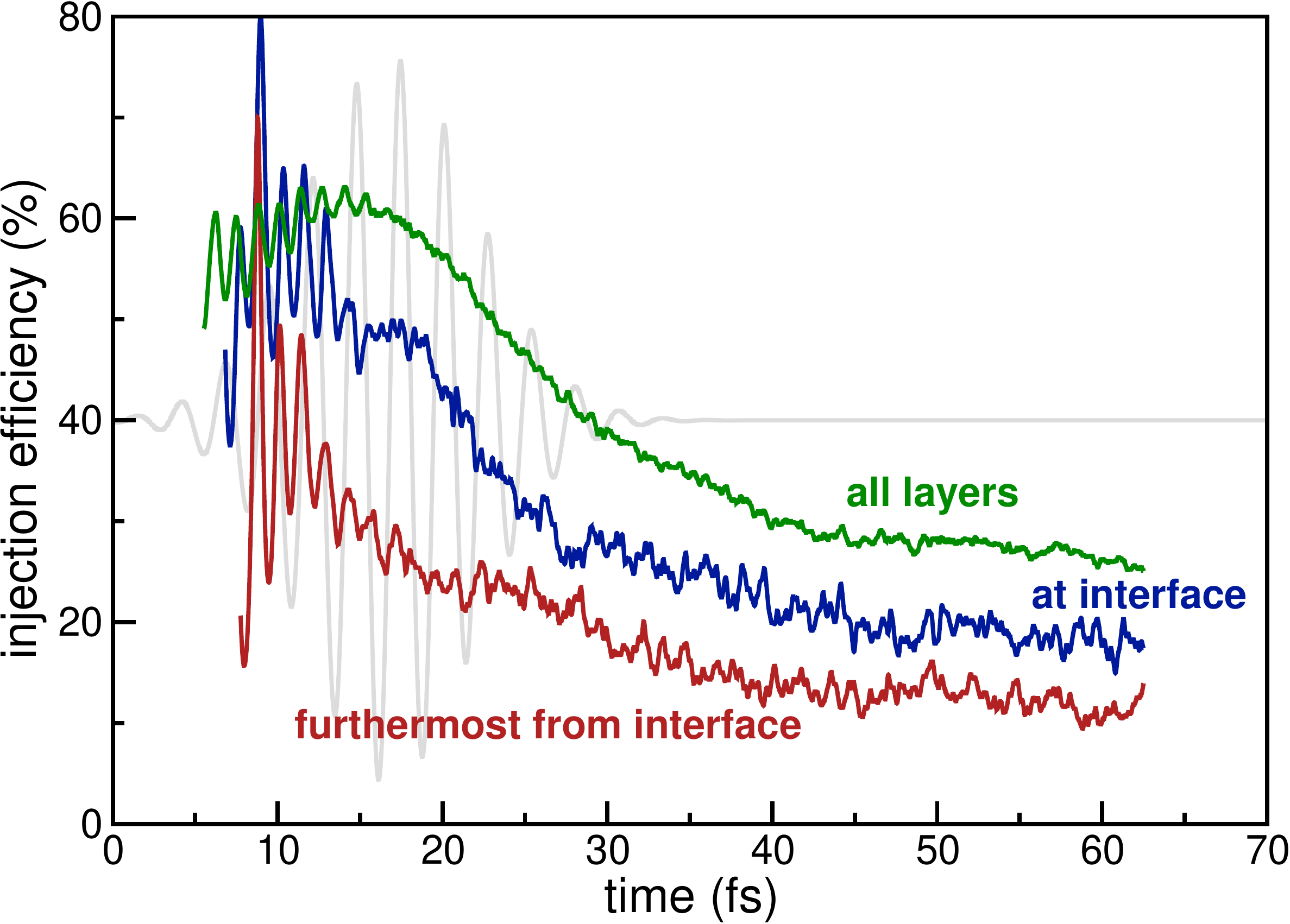}
\caption{Transient spin injection efficiency (given as a percentage, see Eq.~\ref{e:injdef}) for a 3Co/5Cu interface pumped with a laser pulse (vector potential shown in grey) with a central frequency of 1.55~eV, FWHM of 12~fs, and incident fluence of 15~mJ/cm$^2$. The injection efficiency is calculated for the whole structure (all layers), for the two interface layers and for layers furthermost from the interface.} \label{lyr}
\end{figure}

Fig.~\ref{lyr} shows how the SIE depends on the distance from the interface. Given that the initial excitation is a charge transfer excitation between the Co and the Cu layers, the probability of such a transition depends on the overlap of the orbitals, which becomes smaller as the distance from the interface. Hence less spin moment is injected into the furthest Cu layer from the interface. This orbital overlap also implies that the DOS at the interface is quite different from the DOS away from it, which in turn influences the minority back transfer leading to layer dependent SIE. After the initial spin injection, roughly the same decay can be observed in all layers due to similar SOC. A monolayer-resolved experiment which measures the depth dependent transient magnetisation would be able to study such subtle effects in detail. 

\section{Conclusions}

From a quantitative comparison of TDDFT calculations and fs time-resolved SHG experiments, we have shown that the optically induced fs demagnetization of ferromagnet/metal interfaces involves both transient spin injection into the non-ferromagnetic metal and SOC-mediated spin flips. We calculate the attendant spin injection efficiency, finding that an initial SIE of up to 60\% is suppressed within less than 40~fs due to SOC. Choosing metal substrates with lower SOC (e.g. Cu, Al) compared to Au and Pt allows to conserve a SIE of about 30\% even at later times, while the initial SIE is enhanced by choosing interfaces with an electronic DOS that provides sufficient available states for optically induced spin transfer between the ferromagnet and non-ferromagnetic metal layers. Short pump laser pulses (12~fs) are found to increase the initial SIE with increasing fluence, while longer pump pulses (25~fs) at moderate fluence conserve a transient spin injection of nearly 40\%, despite initially lower SIE as compared to the high fluence short pulse regime. 

We can thus design interfaces to increase the available states for optically induced spin injection and tune the pump pulse parameters to enhance the SIE. The present results also point toward the possibility of optimized heterostructures with different metallic layers for enhancing density-of-states effects that promote spin injection and reducing SOC-mediated effects that limit it. 

In order to verify our theoretical predictions, future experimental access to the transient magnetic moments simultaneously in the ferromagnetic and non-ferromagnetic layers will be crucial. In particular, improving fs time-resolved x-ray absorption spectroscopy, including x-ray magnetic circular dichroism, towards higher acquisition efficiency at increased repetition rates is a key experimental goal for the future. In this way one will achieve greater sensitivity to transiently induced magnetic moments in non-ferromagnetic materials than currently possible, an important current development at both large scale facilities such as free electrons lasers and laboratory-based high harmonic generation x-ray sources.

\section{Acknowledgements}

AE, JC and UB were funded by the Deutsche Forschungsgemeinschaft (DFG, German Research Foundation) project-ID 278162697 SFB1242 (projects B01, A07). JKD would like to thank the DFG for funding through project-ID 328545488 TRR227 (project A04). Sharma, Shallcross, AE and JC would like to thank the DFG for funding through SPP 1840 QUTIF Grants No. SH 498/3-1 and ES 492/1-2. PE thanks the DFG for funding through project 438494688. The authors acknowledge the North-German Supercomputing Alliance (HLRN) for providing HPC resources that have contributed to the research results reported in this paper. 

\section{Conflict of Interest} 

The authors declare no conflict of interest. 

\section{Author Contributions} 

AE, JC and UB developed the experimental setup and performed the experimental work. JKD wrote the Elk code, Sharma did the calculations. PE, Shallcross, JKD and Sharma did the analysis of the theoretical results. PE, AE and Sharma wrote the article with input from the remaining authors. 

\section{Data Availability Statement} 

The data that supports the findings of this publication is available from the corresponding author upon request. 

\section{Keywords} 

spintronics, ultrafast spin dynamics, spin injection, spin-orbit coupling, time-dependent density functional theory, second harmonic generation

\bibliography{Ledge}

\providecommand{\latin}[1]{#1}
\makeatletter
\providecommand{\doi}
  {\begingroup\let\do\@makeother\dospecials
  \catcode`\{=1 \catcode`\}=2 \doi@aux}
\providecommand{\doi@aux}[1]{\endgroup\texttt{#1}}
\makeatother
\providecommand*\mcitethebibliography{\thebibliography}
\csname @ifundefined\endcsname{endmcitethebibliography}
  {\let\endmcitethebibliography\endthebibliography}{}
\begin{mcitethebibliography}{33}
\providecommand*\natexlab[1]{#1}
\providecommand*\mciteSetBstSublistMode[1]{}
\providecommand*\mciteSetBstMaxWidthForm[2]{}
\providecommand*\mciteBstWouldAddEndPuncttrue
  {\def\EndOfBibitem{\unskip.}}
\providecommand*\mciteBstWouldAddEndPunctfalse
  {\let\EndOfBibitem\relax}
\providecommand*\mciteSetBstMidEndSepPunct[3]{}
\providecommand*\mciteSetBstSublistLabelBeginEnd[3]{}
\providecommand*\EndOfBibitem{}
\mciteSetBstSublistMode{f}
\mciteSetBstMaxWidthForm{subitem}{(\alph{mcitesubitemcount})}
\mciteSetBstSublistLabelBeginEnd
  {\mcitemaxwidthsubitemform\space}
  {\relax}
  {\relax}

\bibitem[Hirohata \latin{et~al.}(2020)Hirohata, Yamada, Nakatani, Prejbeanu,
  Diény, Pirro, and Hillebrands]{Hirohata2020}
Hirohata,~A.; Yamada,~K.; Nakatani,~Y.; Prejbeanu,~I.-L.; Diény,~B.;
  Pirro,~P.; Hillebrands,~B. Review on spintronics: Principles and device
  applications. \emph{Journal of Magnetism and Magnetic Materials}
  \textbf{2020}, \emph{509}, 166711\relax
\mciteBstWouldAddEndPuncttrue
\mciteSetBstMidEndSepPunct{\mcitedefaultmidpunct}
{\mcitedefaultendpunct}{\mcitedefaultseppunct}\relax
\EndOfBibitem
\bibitem[Jin \latin{et~al.}(2018)Jin, Ma, Karni, Regan, Wang, and
  Heinz]{Jin2018}
Jin,~C.; Ma,~E.~Y.; Karni,~O.; Regan,~E.~C.; Wang,~F.; Heinz,~T.~F. {Ultrafast
  dynamics in van der Waals heterostructures}. \emph{Nat. Nanotechnol.}
  \textbf{2018}, \emph{13}, 994--1003\relax
\mciteBstWouldAddEndPuncttrue
\mciteSetBstMidEndSepPunct{\mcitedefaultmidpunct}
{\mcitedefaultendpunct}{\mcitedefaultseppunct}\relax
\EndOfBibitem
\bibitem[Malinowski \latin{et~al.}(2008)Malinowski, Longa, Rietjens, Paluskar,
  Huijink, Swagten, and Koopmans]{malinowski2008}
Malinowski,~G.; Longa,~F.~D.; Rietjens,~J. H.~H.; Paluskar,~P.~V.; Huijink,~R.;
  Swagten,~H. J.~M.; Koopmans,~B. \emph{Nat. Phys.} \textbf{2008}, \emph{4},
  855\relax
\mciteBstWouldAddEndPuncttrue
\mciteSetBstMidEndSepPunct{\mcitedefaultmidpunct}
{\mcitedefaultendpunct}{\mcitedefaultseppunct}\relax
\EndOfBibitem
\bibitem[Battiato \latin{et~al.}(2010)Battiato, Carva, and
  Oppeneer]{Battiato2010}
Battiato,~M.; Carva,~K.; Oppeneer,~P.~M. {Superdiffusive Spin Transport as a
  Mechanism of Ultrafast Demagnetization}. \emph{Phys. Rev. Lett.}
  \textbf{2010}, \emph{105}, 027203\relax
\mciteBstWouldAddEndPuncttrue
\mciteSetBstMidEndSepPunct{\mcitedefaultmidpunct}
{\mcitedefaultendpunct}{\mcitedefaultseppunct}\relax
\EndOfBibitem
\bibitem[Melnikov \latin{et~al.}(2011)Melnikov, Razdolski, Wehling,
  Papaioannou, Roddatis, Fumagalli, Aktsipetrov, Lichtenstein, and
  Bovensiepen]{melnikov2011}
Melnikov,~A.; Razdolski,~I.; Wehling,~T.~O.; Papaioannou,~E.~T.; Roddatis,~V.;
  Fumagalli,~P.; Aktsipetrov,~O.; Lichtenstein,~A.~I.; Bovensiepen,~U.
  \emph{Phys. Rev. Lett.} \textbf{2011}, \emph{107}, 076601\relax
\mciteBstWouldAddEndPuncttrue
\mciteSetBstMidEndSepPunct{\mcitedefaultmidpunct}
{\mcitedefaultendpunct}{\mcitedefaultseppunct}\relax
\EndOfBibitem
\bibitem[Eschenlohr \latin{et~al.}(2013)Eschenlohr, Battiato, Maldonado,
  Pontius, Kachel, Holldack, Mitzner, Föhlisch, Oppeneer, and
  Stamm]{Eschenlohr2013}
Eschenlohr,~A.; Battiato,~M.; Maldonado,~P.; Pontius,~N.; Kachel,~T.;
  Holldack,~K.; Mitzner,~R.; Föhlisch,~A.; Oppeneer,~P.~M.; Stamm,~C.
  Ultrafast spin transport as key to femtosecond demagnetization. \emph{Nat
  Mater} \textbf{2013}, \emph{12}, 332--336\relax
\mciteBstWouldAddEndPuncttrue
\mciteSetBstMidEndSepPunct{\mcitedefaultmidpunct}
{\mcitedefaultendpunct}{\mcitedefaultseppunct}\relax
\EndOfBibitem
\bibitem[Turgut \latin{et~al.}(2013)Turgut, La-o vorakiat, Shaw, Grychtol,
  Nembach, Rudolf, Adam, Aeschlimann, Schneider, Silva, Murnane, Kapteyn, and
  Mathias]{turgut2013}
Turgut,~E.; La-o vorakiat,~C.; Shaw,~J.~M.; Grychtol,~P.; Nembach,~H.~T.;
  Rudolf,~D.; Adam,~R.; Aeschlimann,~M.; Schneider,~C.~M.; Silva,~T.~J.;
  Murnane,~M.~M.; Kapteyn,~H.~C.; Mathias,~S. Controlling the Competition
  between Optically Induced Ultrafast Spin-Flip Scattering and Spin Transport
  in Magnetic Multilayers. \emph{Phys. Rev. Lett.} \textbf{2013}, \emph{110},
  197201\relax
\mciteBstWouldAddEndPuncttrue
\mciteSetBstMidEndSepPunct{\mcitedefaultmidpunct}
{\mcitedefaultendpunct}{\mcitedefaultseppunct}\relax
\EndOfBibitem
\bibitem[Kampfrath \latin{et~al.}(2013)Kampfrath, Battiato, Maldonado, Eilers,
  N\"otzold, M\"ahrlein, Zbarsky, Freimuth, Mokrousov, Bl\"ugel, Wolf, Radu,
  Oppeneer, and M\"unzenberg]{kampfrath2013}
Kampfrath,~T.; Battiato,~M.; Maldonado,~P.; Eilers,~G.; N\"otzold,~J.;
  M\"ahrlein,~S.; Zbarsky,~V.; Freimuth,~F.; Mokrousov,~Y.; Bl\"ugel,~S.;
  Wolf,~M.; Radu,~I.; Oppeneer,~P.~M.; M\"unzenberg,~M. \emph{Nature Nanotech.}
  \textbf{2013}, \emph{8}, 256\relax
\mciteBstWouldAddEndPuncttrue
\mciteSetBstMidEndSepPunct{\mcitedefaultmidpunct}
{\mcitedefaultendpunct}{\mcitedefaultseppunct}\relax
\EndOfBibitem
\bibitem[Eschenlohr(2020)]{eschenlohr2020}
Eschenlohr,~A. Spin dynamics at interfaces on femtosecond timescales. \emph{J.
  Phys.: Condens. Matter} \textbf{2020}, \emph{33}, 013001\relax
\mciteBstWouldAddEndPuncttrue
\mciteSetBstMidEndSepPunct{\mcitedefaultmidpunct}
{\mcitedefaultendpunct}{\mcitedefaultseppunct}\relax
\EndOfBibitem
\bibitem[Hofherr \latin{et~al.}(2017)Hofherr, Maldonado, Schmitt, Berritta,
  Bierbrauer, Sadashivaiah, Schellekens, Koopmans, Steil, Cinchetti,
  Stadtm\"uller, Oppeneer, Mathias, and Aeschlimann]{hofherr2017}
Hofherr,~M.; Maldonado,~P.; Schmitt,~O.; Berritta,~M.; Bierbrauer,~U.;
  Sadashivaiah,~S.; Schellekens,~A.~J.; Koopmans,~B.; Steil,~D.; Cinchetti,~M.;
  Stadtm\"uller,~B.; Oppeneer,~P.~M.; Mathias,~S.; Aeschlimann,~M. Speed and
  efficiency of femtosecond spin current injection into a nonmagnetic material.
  \emph{Phys. Rev. B} \textbf{2017}, \emph{96}, 100403\relax
\mciteBstWouldAddEndPuncttrue
\mciteSetBstMidEndSepPunct{\mcitedefaultmidpunct}
{\mcitedefaultendpunct}{\mcitedefaultseppunct}\relax
\EndOfBibitem
\bibitem[Chen \latin{et~al.}(2019)Chen, Bovensiepen, Eschenlohr, M\"uller,
  Elliott, Gross, Dewhurst, and Sharma]{chen2019}
Chen,~J.; Bovensiepen,~U.; Eschenlohr,~A.; M\"uller,~T.; Elliott,~P.; Gross,~E.
  K.~U.; Dewhurst,~J.~K.; Sharma,~S. \emph{Phys. Rev. Lett.} \textbf{2019},
  \emph{122}, 067202\relax
\mciteBstWouldAddEndPuncttrue
\mciteSetBstMidEndSepPunct{\mcitedefaultmidpunct}
{\mcitedefaultendpunct}{\mcitedefaultseppunct}\relax
\EndOfBibitem
\bibitem[Dewhurst \latin{et~al.}(2018)Dewhurst, Elliott, Shallcross, Gross, and
  Sharma]{dewhurst2018}
Dewhurst,~J.~K.; Elliott,~P.; Shallcross,~S.; Gross,~E. K.~U.; Sharma,~S.
  Laser-{Induced} {Intersite} {Spin} {Transfer}. \emph{Nano Letters}
  \textbf{2018}, \emph{18}, 1842--1848\relax
\mciteBstWouldAddEndPuncttrue
\mciteSetBstMidEndSepPunct{\mcitedefaultmidpunct}
{\mcitedefaultendpunct}{\mcitedefaultseppunct}\relax
\EndOfBibitem
\bibitem[Alekhin \latin{et~al.}(2017)Alekhin, Razdolski, Ilin, Meyburg,
  Diesing, Roddatis, Rungger, Stamenova, Sanvito, Bovensiepen, and
  Melnikov]{alekhin2017}
Alekhin,~A.; Razdolski,~I.; Ilin,~N.; Meyburg,~J.~P.; Diesing,~D.;
  Roddatis,~V.; Rungger,~I.; Stamenova,~M.; Sanvito,~S.; Bovensiepen,~U.;
  Melnikov,~A. Femtosecond Spin Current Pulses Generated by the Nonthermal
  Spin-Dependent Seebeck Effect and Interacting with Ferromagnets in Spin
  Valves. \emph{Phys. Rev. Lett.} \textbf{2017}, \emph{119}, 017202\relax
\mciteBstWouldAddEndPuncttrue
\mciteSetBstMidEndSepPunct{\mcitedefaultmidpunct}
{\mcitedefaultendpunct}{\mcitedefaultseppunct}\relax
\EndOfBibitem
\bibitem[Siegrist \latin{et~al.}(2019)Siegrist, Gessner, Ossiander, Denker,
  Chang, Schroder, Guggenmos, Cui, Walowski, Martens, Dewhurst, Kleineberg,
  Münzenberg, Sharma, and Schultze]{Siegrist2019}
Siegrist,~F.; Gessner,~J.~A.; Ossiander,~M.; Denker,~C.; Chang,~Y.~P.;
  Schroder,~M.~C.; Guggenmos,~A.; Cui,~Y.; Walowski,~J.; Martens,~U.;
  Dewhurst,~J.~K.; Kleineberg,~U.; Münzenberg,~M.; Sharma,~S.; Schultze,~M.
  Light-wave dynamic control of magnetism. \emph{Nature} \textbf{2019},
  \emph{571}, 240--244\relax
\mciteBstWouldAddEndPuncttrue
\mciteSetBstMidEndSepPunct{\mcitedefaultmidpunct}
{\mcitedefaultendpunct}{\mcitedefaultseppunct}\relax
\EndOfBibitem
\bibitem[Hofherr \latin{et~al.}(2020)Hofherr, H{\"a}user, Dewhurst, Tengdin,
  Sakshath, Nembach, Weber, Shaw, Silva, Kapteyn, Cinchetti, Rethfeld, Murnane,
  Steil, Stadtm{\"u}ller, Sharma, Aeschlimann, and Mathias]{hofherr2020}
Hofherr,~M. \latin{et~al.}  Ultrafast optically induced spin transfer in
  ferromagnetic alloys. \emph{Science Advances} \textbf{2020}, \emph{6}\relax
\mciteBstWouldAddEndPuncttrue
\mciteSetBstMidEndSepPunct{\mcitedefaultmidpunct}
{\mcitedefaultendpunct}{\mcitedefaultseppunct}\relax
\EndOfBibitem
\bibitem[Dewhurst \latin{et~al.}(2020)Dewhurst, Willems, Elliott, Li,
  Schmising, Str\"uber, Engel, Eisebitt, and Sharma]{dewhurst2020}
Dewhurst,~J.~K.; Willems,~F.; Elliott,~P.; Li,~Q.~Z.; Schmising,~C. v.~K.;
  Str\"uber,~C.; Engel,~D.~W.; Eisebitt,~S.; Sharma,~S. Element Specificity of
  Transient Extreme Ultraviolet Magnetic Dichroism. \emph{Phys. Rev. Lett.}
  \textbf{2020}, \emph{124}, 077203\relax
\mciteBstWouldAddEndPuncttrue
\mciteSetBstMidEndSepPunct{\mcitedefaultmidpunct}
{\mcitedefaultendpunct}{\mcitedefaultseppunct}\relax
\EndOfBibitem
\bibitem[Sharma \latin{et~al.}(2022)Sharma, Shallcross, Elliott, Eisebitt,
  v.~Korff~Schmising, and Dewhurst]{sharma22}
Sharma,~S.; Shallcross,~S.; Elliott,~P.; Eisebitt,~S.; v.~Korff~Schmising,~C.;
  Dewhurst,~J.~K. Computational analysis of transient XMCD sum rules for laser
  pumped systems: When do they fail? \emph{Appl. Phys. Lett.} \textbf{2022},
  \emph{120}, 062409\relax
\mciteBstWouldAddEndPuncttrue
\mciteSetBstMidEndSepPunct{\mcitedefaultmidpunct}
{\mcitedefaultendpunct}{\mcitedefaultseppunct}\relax
\EndOfBibitem
\bibitem[Dewhurst \latin{et~al.}(2018)Dewhurst, Shallcross, Gross, and
  Sharma]{dewhurst2018a}
Dewhurst,~J.~K.; Shallcross,~S.; Gross,~E. K.~U.; Sharma,~S.
  Substrate-Controlled Ultrafast Spin Injection and Demagnetization.
  \emph{Phys. Rev. Applied} \textbf{2018}, \emph{10}, 044065\relax
\mciteBstWouldAddEndPuncttrue
\mciteSetBstMidEndSepPunct{\mcitedefaultmidpunct}
{\mcitedefaultendpunct}{\mcitedefaultseppunct}\relax
\EndOfBibitem
\bibitem[Slonczewski(1996)]{SLONCZEWSKI1996}
Slonczewski,~J. Current-driven excitation of magnetic multilayers.
  \emph{Journal of Magnetism and Magnetic Materials} \textbf{1996}, \emph{159},
  L1--L7\relax
\mciteBstWouldAddEndPuncttrue
\mciteSetBstMidEndSepPunct{\mcitedefaultmidpunct}
{\mcitedefaultendpunct}{\mcitedefaultseppunct}\relax
\EndOfBibitem
\bibitem[Berger(1996)]{Berger1996}
Berger,~L. Emission of spin waves by a magnetic multilayer traversed by a
  current. \emph{Phys. Rev. B} \textbf{1996}, \emph{54}, 9353--9358\relax
\mciteBstWouldAddEndPuncttrue
\mciteSetBstMidEndSepPunct{\mcitedefaultmidpunct}
{\mcitedefaultendpunct}{\mcitedefaultseppunct}\relax
\EndOfBibitem
\bibitem[Runge and Gross(1984)Runge, and Gross]{RG1984}
Runge,~E.; Gross,~E. K.~U. Density-Functional Theory for Time-Dependent
  Systems. \emph{Phys. Rev. Lett.} \textbf{1984}, \emph{52}, 997--1000\relax
\mciteBstWouldAddEndPuncttrue
\mciteSetBstMidEndSepPunct{\mcitedefaultmidpunct}
{\mcitedefaultendpunct}{\mcitedefaultseppunct}\relax
\EndOfBibitem
\bibitem[Sharma \latin{et~al.}(2014)Sharma, Dewhurst, and Gross]{my-book}
Sharma,~S.; Dewhurst,~J.~K.; Gross,~E. K.~U. In \emph{First Principles
  Approaches to Spectroscopic Properties of Complex Materials};
  Di~Valentin,~C., Botti,~S., Cococcioni,~M., Eds.; Springer Berlin Heidelberg:
  Berlin, Heidelberg, 2014; pp 235--257\relax
\mciteBstWouldAddEndPuncttrue
\mciteSetBstMidEndSepPunct{\mcitedefaultmidpunct}
{\mcitedefaultendpunct}{\mcitedefaultseppunct}\relax
\EndOfBibitem
\bibitem[Krieger \latin{et~al.}(2015)Krieger, Dewhurst, Elliott, Sharma, and
  Gross]{krieger2015}
Krieger,~K.; Dewhurst,~J.~K.; Elliott,~P.; Sharma,~S.; Gross,~E. K.~U.
  Laser-{Induced} {Demagnetization} at {Ultrashort} {Time} {Scales}:
  {Predictions} of {TDDFT}. \emph{Journal of Chemical Theory and Computation}
  \textbf{2015}, \emph{11}, 4870--4874\relax
\mciteBstWouldAddEndPuncttrue
\mciteSetBstMidEndSepPunct{\mcitedefaultmidpunct}
{\mcitedefaultendpunct}{\mcitedefaultseppunct}\relax
\EndOfBibitem
\bibitem[Dewhurst \latin{et~al.}(2016)Dewhurst, Krieger, Sharma, and
  Gross]{dewhurst2016}
Dewhurst,~J.~K.; Krieger,~K.; Sharma,~S.; Gross,~E. K.~U. An efficient
  algorithm for time propagation as applied to linearized augmented plane wave
  method. \emph{Computer Physics Communications} \textbf{2016}, \emph{209},
  92--95\relax
\mciteBstWouldAddEndPuncttrue
\mciteSetBstMidEndSepPunct{\mcitedefaultmidpunct}
{\mcitedefaultendpunct}{\mcitedefaultseppunct}\relax
\EndOfBibitem
\bibitem[Singh(1994)]{singh}
Singh,~D.~J. \emph{Planewaves Pseudopotentials and the LAPW Method}; Kluwer
  Academic Publishers, Boston, 1994\relax
\mciteBstWouldAddEndPuncttrue
\mciteSetBstMidEndSepPunct{\mcitedefaultmidpunct}
{\mcitedefaultendpunct}{\mcitedefaultseppunct}\relax
\EndOfBibitem
\bibitem[Dewhurst \latin{et~al.}(Jan. 14 {\bf 2018})Dewhurst, Sharma, and
  et~al.]{elk}
Dewhurst,~J.~K.; Sharma,~S.; et~al., Jan. 14 {\bf 2018};
  \url{elk.sourceforge.net}\relax
\mciteBstWouldAddEndPuncttrue
\mciteSetBstMidEndSepPunct{\mcitedefaultmidpunct}
{\mcitedefaultendpunct}{\mcitedefaultseppunct}\relax
\EndOfBibitem
\bibitem[Weber \latin{et~al.}(1996)Weber, Bischof, Allenspach, Back,
  Fassbender, May, Schirmer, Jungblut, G\"untherodt, and
  Hillebrands]{weber1996}
Weber,~W.; Bischof,~A.; Allenspach,~R.; Back,~C.~H.; Fassbender,~J.; May,~U.;
  Schirmer,~B.; Jungblut,~R.; G\"untherodt,~G.; Hillebrands,~B. \emph{Phys.
  Rev. B} \textbf{1996}, \emph{54}, 4075\relax
\mciteBstWouldAddEndPuncttrue
\mciteSetBstMidEndSepPunct{\mcitedefaultmidpunct}
{\mcitedefaultendpunct}{\mcitedefaultseppunct}\relax
\EndOfBibitem
\bibitem[J\"ahnke \latin{et~al.}(1999)J\"ahnke, Conrad, G\"udde, and
  Matthias]{jaehnke1999}
J\"ahnke,~V.; Conrad,~U.; G\"udde,~J.; Matthias,~E. \emph{Appl. Phys. B}
  \textbf{1999}, \emph{68}, 485\relax
\mciteBstWouldAddEndPuncttrue
\mciteSetBstMidEndSepPunct{\mcitedefaultmidpunct}
{\mcitedefaultendpunct}{\mcitedefaultseppunct}\relax
\EndOfBibitem
\bibitem[G\"udde \latin{et~al.}(1999)G\"udde, Conrad, J\"ahnke, Hohlfeld, and
  Matthias]{guedde1999}
G\"udde,~J.; Conrad,~U.; J\"ahnke,~V.; Hohlfeld,~J.; Matthias,~E. \emph{Phys.
  Rev. B} \textbf{1999}, \emph{59}, R6608\relax
\mciteBstWouldAddEndPuncttrue
\mciteSetBstMidEndSepPunct{\mcitedefaultmidpunct}
{\mcitedefaultendpunct}{\mcitedefaultseppunct}\relax
\EndOfBibitem
\bibitem[Chen \latin{et~al.}(2017)Chen, Wieczorek, Eschenlohr, Xiao,
  Tarasevitch, and Bovensiepen]{chen2017}
Chen,~J.; Wieczorek,~J.; Eschenlohr,~A.; Xiao,~S.; Tarasevitch,~A.;
  Bovensiepen,~U. \emph{Appl. Phys. Lett.} \textbf{2017}, \emph{110},
  092407\relax
\mciteBstWouldAddEndPuncttrue
\mciteSetBstMidEndSepPunct{\mcitedefaultmidpunct}
{\mcitedefaultendpunct}{\mcitedefaultseppunct}\relax
\EndOfBibitem
\bibitem[Conrad \latin{et~al.}(2001)Conrad, G\"udde, J\"ahnke, and
  Matthias]{conrad2001}
Conrad,~U.; G\"udde,~J.; J\"ahnke,~V.; Matthias,~E. \emph{Phys. Rev. B}
  \textbf{2001}, \emph{63}, 144417\relax
\mciteBstWouldAddEndPuncttrue
\mciteSetBstMidEndSepPunct{\mcitedefaultmidpunct}
{\mcitedefaultendpunct}{\mcitedefaultseppunct}\relax
\EndOfBibitem
\bibitem[Krieger \latin{et~al.}(2015)Krieger, Dewhurst, Elliott, Sharma, and
  Gross]{KDES15}
Krieger,~K.; Dewhurst,~J.~K.; Elliott,~P.; Sharma,~S.; Gross,~E. K.~U. \emph{J.
  Chem. Theory Comput.} \textbf{2015}, \emph{11}, 4870--4874\relax
\mciteBstWouldAddEndPuncttrue
\mciteSetBstMidEndSepPunct{\mcitedefaultmidpunct}
{\mcitedefaultendpunct}{\mcitedefaultseppunct}\relax
\EndOfBibitem
\end{mcitethebibliography}

\end{document}